\documentstyle[preprint,aps,12pt,epsf]{revtex}
\tightenlines
\begin{document}
\draft
\preprint{\vtop{{\hbox{YITP-99-65}\vskip-0pt
}}}
\title{Non-factorizable contributions in $B$ decays
\footnote{Contribution to the Third International Conference 
on $B$ Physics and $CP$ Violation, Taipei, Taiwan, December 3 -- 7, 
1999.}
}
\author{K. Terasaki\\ Yukawa Institute for Theoretical Physics,\\
Kyoto University, Kyoto 606-8502, Japan
}
\maketitle
\thispagestyle{empty}
\begin{abstract}

First of all, $\bar B \rightarrow D\pi$ and $D^*\pi$ decays are 
studied phenomenologically and possible lower bounds of the branching 
ratios, 
$B(\bar B^0\rightarrow D^0\pi^0)$ and 
$B(\bar B^0\rightarrow D^{*0}\pi^0)$, 
are estimated from existing experimental data on branching ratios 
for the $\bar B \rightarrow D\pi$ and $D^*\pi$ decays. 

Then, 
$\bar B \rightarrow D\pi$, $D^*\pi$, $J/\psi\bar K$ and $J/\psi\pi$ 
decays are studied by decomposing their amplitude into a sum of 
factorizable and non-factorizable ones. The former is estimated by
using the naive factorization while the latter is calculated by using 
a hard pion approximation in the infinite momentum frame. 

The result is compared with the above
phenomenological branching ratios (and their observed ones). 
The non-factorizable amplitude is rather small in the color favored 
$\bar B \rightarrow D\pi$ and $D^*\pi$ decays but can still
efficiently interfere with the main (factorized) amplitude. 
In the color suppressed $\bar B \rightarrow J/\psi\bar K$ and
$J/\psi\pi$ decays, non-factorizable contribution is more important. 
A sum of the factorized and non-factorizable amplitudes can improve 
the result from the factorization, although the amplitudes for the 
color suppressed 
$\bar B^0 \rightarrow D^0\pi^0$, $D^{*0}\pi^0$ and 
$\bar B\rightarrow J/\psi \bar K$, $J/\psi\pi$ 
decays still include ambiguities arising from uncertainties of form 
factors involved. 

\end{abstract}

\vskip 0.5cm
\newpage
\section{Introduction}

Nonleptonic weak decays of charm and $B$ mesons have been studied 
extensively\cite{BSW,NRSX} by using the so-called factorization (or 
vacuum insertion) prescription\cite{Oneda-Wakasa}. It has been
supported by two independent arguments. One is the large $N_c$
(color degree of freedom) argument\cite{Large-N} that the 
factorizable amplitude which is given by the leading terms in the 
large $N_c$ expansion dominates in hadronic weak decays. 
The other is that it can be a good approximation under a certain 
kinematical condition\cite{DG}, i.e., a heavy quark decays into 
another heavy quark plus a pair of light quark and anti-quark which 
are emitted colinearly with sufficiently high energies, for example, 
like $b \rightarrow c\,+\, (\bar ud)_1$, where $(\bar ud)_1$ denotes 
a color singlet $(\bar ud)$ pair. 

We, first, consider two body decays of charm mesons to check whether
the large $N_c$ argument works well or not in  hadronic weak 
interactions, since the large $N_c$ argument is independent of
flavors, i.e., if it does not work in charm decays, it does not work 
also in $B$ decays. A naive application of the factorization  
prescription to charm decay amplitudes leads to the color 
suppression [suppression of color mismatched decays, 
$D^0 \rightarrow \bar K^0\pi^0,\,\bar K^{*0}\pi^0$, etc., 
described by 
$c \rightarrow (s\bar d)_1 \,+\, u$]. 
It means, for example, that the amplitude for the decays of charm 
mesons into isospin $I={1\over 2}\,\,(\bar K\pi)$ final states is 
approximately cancelled by the one into $I={3\over 2}$ final states 
and hence the phases of these amplitudes are nearly equal to each 
other. Therefore the factorized amplitudes for two body decays of 
charm mesons should be approximately real except for the overall 
phase. However the observed decay rates for these decays are not 
always suppressed and the amplitudes for $D \rightarrow \bar K\pi$  
and $\bar K^*\pi$ decays have large phase differences between the 
amplitudes for decays into the $I={1\over 2}$ and $I={3\over 2}$ 
final states\cite{BHP}. To get rid of this problem, the factorization
has been implemented by taking account for final state interactions. 
However, amplitudes with final state interactions are given by 
non-leading terms in the large $N_c$ expansion. Therefore the large 
$N_c$ argument does not work well in charm decays and hence also in 
$B$ decays. 

It appears that, in $\bar B\rightarrow D\pi$ [and $D^*\pi$] decays, 
the color suppression works well and the phase differences between 
amplitudes for decays into $I={1\over 2}$ and $I={3\over 2}$ final 
states are small. To see it explicitly, we parameterize the 
amplitudes for these decays as 
\begin{eqnarray}
&&A(\bar B^0\, \rightarrow D^{[*]+}\pi^-) 
= \,\,\,\,\sqrt{{1 \over 3}}A^{[*]}_{3}e^{i\delta_{3}^{[*]}} 
+ \sqrt{{2 \over 3}}A^{[*]}_{1}e^{i\delta_{1}^{[*]}},  
                                               \label{eq:amp-pm}\\
&&A(\bar B^0\, \rightarrow D^{[*]0}\,\pi^0\,) 
= -\sqrt{{2 \over 3}}A^{[*]}_{3}e^{i\delta_{3}^{[*]}} 
+  \sqrt{{1 \over 3}}A^{[*]}_{1}e^{i\delta_{1}^{[*]}},  
                                               \label{eq:amp-zz}\\
&&A(B^- \rightarrow  D^{[*]0}\pi^-) 
= \quad\sqrt{3}A^{[*]}_{3}e^{i\delta_{3}^{[*]}},            
                                                \label{eq:amp-zm}
\end{eqnarray}
where $A^{[*]}_{2I}$'s and $\delta_{2I}^{[*]}$'s are isospin eigen 
amplitudes for the $\bar B\rightarrow D^{[*]}\pi$ decays and their 
phases, respectively. Taking positive values of the ratio of isospin 
eigen amplitudes, $r^{[*]}=A_{3}^{[*]}/A^{[*]}_{1}$, we obtain 
\begin{equation}
\cos(\delta^{[*]}) 
= \Bigl({9R_{0-}^{[*]} - 1\over 4}\Bigr)r^{[*]} - {1\over r^{[*]}}, 
                                           \label{eq:phase-diff}
\end{equation}
where 
\begin{equation}
\delta^{[*]} = \delta_1^{[*]} -\delta_3^{[*]} \quad {\rm and} \quad 
r^{[*]} = \sqrt{1\over
                      3R_{00}^{[*]}R_{0-}^{[*]}+3R_{0-}^{[*]}-1}\,\,.     
                                              \label{eq:ratio-amps}
\end{equation} 
Here $R_{0-}^{[*]}$ and $R_{00}^{[*]}$ are ratios of decay rates, 
\begin{equation}
R_{0-}^{[*]} = {\Gamma(\bar B^0\rightarrow D^{[*]+}\pi^-) 
\over\Gamma(B^-\rightarrow D^{[*]0}\pi^-)} 
\quad {\rm and}\quad
R_{00}^{[*]} = {\Gamma(\bar B^0\rightarrow D^{[*]0}\pi^0) 
\over \Gamma(\bar B^0\rightarrow D^{[*]+}\pi^-)}.
                                           \label{eq:ratio-rates}
\end{equation}
Values of $R_{0-}^{[*]}$ and $R_{00}^{[*]}$ can be estimated 
phenomenologically from the experimental data~\cite{PDG98} on  
branching ratios for $\bar B\rightarrow D^{[*]}\pi$ decays in 
Table~II as 
\begin{eqnarray}
&&R_{0-} = 0.58 \pm 0.10, \quad R_{00} < 0.04,     \\
                                         \label{eq:value-of-R} 
&& R_{0-}^{*} = 0.61 \pm 0.08, \quad R_{00}^{*} < 0.16. 
                                         \label{eq:value-of-RS}
\end{eqnarray}
However, these values of $R_{0-}$ and $R_{00}$ [$R_{0-}^{*}$ and
$R_{00}^{*}$] are not always compatible with each other. If all the 
above values of  $R^{[*]}$'s are accepted, then the right-hand-side 
(r.h.s.) in Eq.(\ref{eq:phase-diff}) is not always less than unity. 
It is satisfied in more restricted regions of $R^{[*]}$, i.e., 
approximately, 
$0.04 \gtrsim R_{00} \gtrsim 0.02$,  
$0.68 \gtrsim  R_{0-} \gtrsim 0.61$  
for the $\bar B\rightarrow D\pi$ decays and 
$0.16 \gtrsim R^*_{00} \gtrsim 0.02$, 
$0.69 \gtrsim R^*_{0-}\gtrsim 0.53$ 
for the $\bar B\rightarrow D^*\pi$ decays. 
These values of $R^{[*]}$ lead approximately to the following 
phenomenological branching ratios, 
$0.34 \gtrsim B(\bar B^0 \rightarrow D^+\pi^-)_{\rm ph}\gtrsim 0.28$ 
and 
$0.012 \gtrsim B(\bar B^0 \rightarrow D^0\pi^0)_{\rm ph} \gtrsim
0.007$, 
and 
$0.044 \gtrsim B(\bar B^0 \rightarrow D^{*0}\pi^0)_{\rm ph}
\gtrsim 0.004$, 
when the experimental data, 
$B(B^- \rightarrow D^0\pi^-)_{\rm expt}= 0.53 \pm 0.05$ and 
$B(B^- \rightarrow D^{*0}\pi^-)_{\rm expt}= 0.46 \pm 0.04$, 
are fixed. Here we put 
$B(\bar B^0 \rightarrow D^{*+}\pi^-)_{\rm ph}
=B(\bar B^0 \rightarrow D^{*+}\pi^-)_{\rm expt}$ 
since $\cos(\delta^{*}) \leq 1$ is satisfied for all the
experimentally allowed values of $R_{0-}^*$. The allowed values of 
$\cos(\delta)$ are limited within a narrow region, 
$0.96 \lesssim \,\cos(\delta)\,\,\, \leq \,\,1$ 
in which $\cos(\delta)$ is very close to unity while $\cos(\delta^*)$ 
is a little more mildly restricted (at the present stage) compared 
with the above $\cos(\delta)$, i.e., approximately, 
$0.70 \lesssim \cos(\delta^{*}) \leq\,\, 1 $. 
Therefore, the color suppression works well, at least, in 
$\bar B \rightarrow D\pi$ decays and the phase difference $\delta$ is 
very small. In this way, it will be understood that the factorized
amplitudes are dominant only in some specific decays like the 
$\bar B\rightarrow D\pi$ decays but it is still a little ambiguous in 
the $\bar B \rightarrow D^*\pi$ decays. 

In this article, we study $\bar B \rightarrow D\pi$, $D^*\pi$, 
$J/\psi\bar K$ and $J/\psi\pi$ decays. In the next section, we will 
present our basic assumption and review briefly the effective weak 
Hamiltonian. In Sec.~III, the $\bar B \rightarrow D\pi$ and $D^*\pi$ 
decays will be studied by decomposing their amplitude into a sum of 
factorizable and non-factorizable ones. The former will be estimated 
by using the naive factorization while the latter is calculated by 
using a hard pion approximation in the infinite momentum frame. 
In Sec.~IV, the color suppressed decays, 
$\bar B \rightarrow J/\psi\bar K$ and 
$B^- \rightarrow J/\psi\pi^-$, 
will be investigated in the same way. A brief summary will be given 
in the final section. 

\section{Basic Assumption and Effective Weak Hamiltonian}

Our starting point to study nonleptonic weak processes is to decompose 
their amplitude into a sum of {\it factorizable} and 
{\it non-factorizable} ones~\cite{Terasaki-B}. Therefore, the
effective weak Hamiltonian should be divided into the corresponding 
parts, 
\begin{equation}
H_w = (H_w)_{\rm FA} + (H_w)_{\rm NF},    
                                                \label{eq:HW-dev}
\end{equation}
where $(H_w)_{\rm FA}$ and $(H_w)_{\rm NF}$ are responsible for
factorizable and non-factorizable amplitudes, respectively. The 
factorizable amplitude is estimated by using the naive 
factorization\cite{BSW,Oneda-Wakasa}. Then, assuming that the 
non-factorizable amplitude is dominated by dynamical contributions 
of various hadrons\cite{hadron-dynamics} and using a hard pion
approximation in the infinite momentum frame 
(IMF)\cite{hard pion,suppl}, we estimate the  non-factorizable 
amplitudes. The hard pion amplitude will be given by {\it asymptotic} 
matrix elements of $(H_w)_{\rm NF}$ [matrix elements  of 
$(H_w)_{\rm NF}$ taken between single hadron states with infinite 
momentum]. 

Before we study amplitudes for $B$ decays, we review briefly the 
$|\Delta B|=1$ effective weak Hamiltonian. Its main part is usually 
written in the  form 
\begin{equation}
H_w = {G_F \over \sqrt{2}}V_{ud}V_{bc}\Bigl\{c_1O_1 
                                      + c_2O_2 \Bigr\} +  h.c.,  
                                                    \label{eq:HW}
\end{equation}
where the four quark operators $O_1$ and $O_2$ are given by 
products of color singlet left-handed currents, 
\begin{equation}
O_1 = :(\bar cb)_{V-A}(\bar du)_{V-A}: \quad {\rm and} \quad 
O_2 = :(\bar cu)_{V-A}(\bar db)_{V-A}: .          \label{eq:FQO}
\end{equation}
$V_{ij}$ denotes a CKM matrix element\cite{CKM} which is taken to 
be real since CP invariance is always assumed in this paper. 

When we calculate the factorizable amplitudes for the 
$\bar B\rightarrow D^{[*]}\pi$ decays later, we use, as usual, the 
so-called BSW Hamiltonian\cite{BSW,NRSX} 
\begin{equation}
H_w^{BSW} = {G_F \over \sqrt{2}}V_{ud}V_{cb}
       \Bigl\{a_1O_1 + a_2O_2 \Bigr\} + h.c.   \label{eq:HW-BSW}
\end{equation}
which can be obtained from Eq.(\ref{eq:HW}) by using the Fierz 
reordering. The operators $O_1$ and $O_2$ in Eq.(\ref{eq:HW-BSW}) 
should be no longer Fierz reordered.  We, therefore, replace 
$(H_w)_{\rm FA}$ by $H_w^{BSW}$. The coefficients $a_1$ and $a_2$ 
are given by 
\begin{equation}
a_1 = c_1 + {c_2 \over N_c},\quad a_2 = c_2 + {c_1 \over N_c},
                                               \label{eq:coef-BSW}
\end{equation}
where $N_c$ is the color degree of freedom. 

When $H_w^{BSW}$ is obtained, an extra term which is given by a 
color singlet sum of products of colored currents, 
\begin{equation}
\tilde H_w = {G_F \over \sqrt{2}}V_{ud}V_{cb}
\Bigl\{c_2\tilde O_1 + c_1\tilde O_2 \Bigr\} + h.c., 
                                          \label{eq:HW-Non-f}
\end{equation}
comes out, where 
\begin{equation}
\tilde O_1 = 2\sum_a:(\bar ct^ab)_{V-A}(\bar dt^au)_{V-A}: 
\quad {\rm and} \quad 
\tilde O_2 = 2\sum_a:(\bar ct^au)_{V-A}(\bar dt^ab)_{V-A}:         
                                             \label{eq:FQO-extra}
\end{equation}
with the generators $t^a$ of the color $SU_c(N_c)$ symmetry. To
describe physical amplitudes for $B$ decays by matrix elements of
$\tilde H_w$, soft gluon(s) should be exchanged between quarks which 
belong to different meson states. Therefore, amplitudes given by 
$\tilde H_w$ correspond to non-leading terms in the large $N_c$ 
expansion and not factorizable so that $(H_w)_{\rm NF}$ is now 
replaced by $\tilde H_w$. 

\section{$\bar B \rightarrow D\pi$ and $D^*\pi$ decays}

The factorization prescription in the BSW scheme leads to the
following factorized amplitude, for example, for the 
$B^-(p) \rightarrow D^0(p')\pi^-(q)$ decay, 
\begin{eqnarray}
&& M_{\rm FA}(B^-(p) \rightarrow D^0(p')\pi^-(q))   \nonumber\\
&&\hspace{20mm}
= {G_F \over \sqrt{2}}V_{cb}V_{ud}
\Bigl\{ 
a_1\langle \pi^-(q)|(\bar du)_{V-A}|0\rangle 
\langle D^0(p')|(\bar cb)_{V-A}|B^-(p) \rangle      \nonumber\\   
&&\hspace{45mm} 
+ a_2\langle D^0(p')|(\bar cu)_{V-A}|0\rangle
\langle \pi^-(q)|(\bar db)_{V-A}|B^-(p) \rangle
\Bigr\}.
                                                    \label{eq:FACT}
\end{eqnarray}
Factorizable amplitudes for the other $\bar B \rightarrow D\pi$ and
$D^*\pi$ decays also can be calculated in the same way. To evaluate 
these amplitudes, we use the following parameterization of matrix 
elements of currents in Ref.\cite{NRSX},  
\begin{equation}
       \langle \pi(q)|A_\mu^{}|0 \rangle = -if_{\pi}q_\mu, 
                                                   \label{eq:PCAC}
\end{equation}
\begin{eqnarray}
&&\langle D(p')|V_\mu^{}|\bar B(p)\rangle 
= \Biggl\{(p+p')_\mu - {m_B^2 - m_D^2 \over q^2}q_\mu\Biggr\}
F_1(q^2) 
        + {m_B^2 - m_D^2 \over q^2}q_\mu F_0(q^2),  \label{eq:DB}
\\ &&\langle D^*(p')|A_\mu^{}|\bar B(p)\rangle 
= \Biggl\{
(m_B + m_{D^*})\epsilon_\mu^*(p')A_1(q^2) 
- {\epsilon^*(p')\cdot q \over m_B + m_{D^*}}(p + p')_\mu A_2(q^2)
\nonumber \\
&& \hskip 6cm
- 2m_{D^*}{\epsilon^*\cdot q \over q^2}q_\mu A_3(q^2)
\Biggr\}
+ 2m_{D^*}{\epsilon^*\cdot q \over q^2}q_\mu A_0(q^2), 
                                                    \label{eq:D^*B}
\end{eqnarray}
where $q = p - p'$ and the form factors satisfy 
\begin{eqnarray}
&& A_3(q^2) = {m_B + m_{D^*} \over 2m_{D^*}}A_1(q^2) 
- {m_B - m_{D^*} \over 2m_{D^*}}A_2(q^2),         \label{eq:FFA}\\
&& 
F_1(0) = F_0(0), \qquad A_3(0) = A_0(0).           \label{eq:FF0}
\end{eqnarray}
To get rid of useless imaginary unit except for the overall phase 
in the amplitude, however, we adopt the following parameterization
of  matrix element of vector current\cite{HY}, 
\begin{equation}
\langle V(p')|V_\mu^{}|0\rangle 
        = -if_{V}m_{V}\epsilon_{\mu}^*(p').      \label{eq:Lvector}
\end{equation}
As stressed in Ref.\cite{HY}, the above matrix element of vector 
current can be treated in parallel to those of axial vector 
currents in Eq.(\ref{eq:PCAC}) in the infinite momentum frame
(IMF).  Using these expressions of current matrix elements, we
obtain  the factorized amplitudes for $\bar B \rightarrow D\pi$ and 
$D^*\pi$ decays in Table~I, where we have put $m_\pi^2=0$. 

Before we evaluate numerically the factorized amplitudes, we study 
non-factorizale amplitudes for $\bar B \rightarrow D\pi$ and 
$D^*\pi$ decays. To this, we assume that the non-factorizable
amplitudes are  dominated by dynamical contributions of various
hadron states. Then they can be estimated by using a hard pion
technique in the IMF; i.e., 
${\bf p} \rightarrow \infty$\cite{hard pion,suppl}. 
It is an innovation of the old soft pion technique\cite{soft pion}. 
In our hard pion approximation, the non-factorizable amplitude for 
the $\bar B({p}) \rightarrow D^{[*]}({p'})\pi({q})$ decay is given by 
\begin{equation}
M_{\rm NF}(\bar B \rightarrow D^{[*]}\pi) 
\simeq M_{\rm ETC}(\bar B \rightarrow D^{[*]}\pi) 
+ M_{\rm S}(\bar B \rightarrow D^{[*]}\pi).        
                                                      \label{eq:HP}
\end{equation} 
The equal-time commutator term ($M_{\rm ETC}$) and the surface term
($M_{\rm S}$) are given by 
\newpage
\begin{center}
\begin{quote}
{Table~I. Factorized amplitudes for $\bar B \rightarrow D\pi$ and 
$D^*\pi$ decays where $m_\pi^2=0$. The CKM matrix elements are
factored out.}
\end{quote}
\vspace{0.5cm}

\begin{tabular}
{l|l}
\hline
$\quad\,\,${\rm Decay}
&\hskip 5cm {$\quad A_{\rm FA}\,$}
\\
\hline 
$\bar B^0 \rightarrow D^{+}\pi^-$
& $\,\quad i
   {G_F \over\sqrt{2}}a_1f_\pi(m_B^2 - m_D^2)F_0^{DB}(0) 
\Bigl[
1 - \Bigl({a_2 \over a_1}\Bigr)\Bigl({f_B \over f_\pi}\Bigr)
\Bigl({m_D^2 \over m_B^2 - m_D^2}\Bigr)
{F_0^{D\pi}(m_B^2) \over F_0^{DB}(0)}
\Bigr]$ 
\\
$\bar B^0 \rightarrow D^{0}\pi^0$
& $\,-\,i
   {G_F \over {2}}a_2f_Dm_B^2F_0^{\pi B}(m_D^2) 
\Bigl[
1 + \Bigl({f_B \over f_D}\Bigr)
\Bigl({m_D^2 \over m_B^2}\Bigr)
{F_0^{D\pi}(m_B^2) \over F_0^{\pi B}(m_D^2)}
\Bigr]
$
\\
$B^- \rightarrow D^0\pi^-$
& $\,\quad i
   {G_F \over\sqrt{2}}a_1f_\pi(m_B^2 - m_D^2)F_0^{DB}(0) 
\Bigl[
1 + \Bigl({a_2 \over a_1}\Bigr)\Bigl({f_D \over f_\pi}\Bigr)
\Bigl({m_B^2 \over m_B^2 - m_D^2}\Bigr)
{F_0^{\pi B}(m_D^2) \over F_0^{DB}(0)}
\Bigr]
$ 
\\
$\bar B^0 \rightarrow D^{*+}\pi^-$
& $\,-\,i
   {G_F \over\sqrt{2}}a_1f_\pi A_0^{D^*B}(0) 
\Bigl[
1 - \Bigl({a_2 \over a_1}\Bigr)\Bigl({f_B \over f_\pi}\Bigr)
{A_0^{D^*\pi}(m_B^2) \over A_0^{D^*B}(0)}
\Bigr]2m_{D^*}\epsilon^*(p')\cdot p$
\\
$\bar B^0 \rightarrow D^{*0}\pi^0$
& $\,\quad i
   {G_F \over {2}}a_2f_{D^*}F_1^{\pi B}(m_D^{*2}) 
\Bigl[
1 + \Bigl({f_B \over f_{D^*}}\Bigr)
{A_0^{D^*\pi}(m_B^2) \over F_1^{\pi B}(m_D^{*2})}
\Bigr]2m_{D^*}\epsilon^*(p')\cdot p$
\\
$B^- \rightarrow D^{*0}\pi^-$
& $\,-\,i
   {G_F \over\sqrt{2}}a_1f_\pi A_0^{D^*B}(0) 
\Bigl[
1 + \Bigl({a_2 \over a_1}\Bigr)\Bigl({f_{D^*} \over f_\pi}\Bigr)
{F_1^{\pi B}(m_D^{*2}) \over A_0^{D^*B}(0)}
\Bigr]2m_{D^*}\epsilon^*(p')\cdot p$ 
\\
\hline
\end{tabular}

\end{center}
\vspace{0.5cm}
\begin{equation}
M_{\rm ETC}(\bar B \rightarrow D^{[*]}\pi) 
= {i \over f_{\pi}}
        \langle{D^{[*]}|[V_{\bar \pi}, \tilde H_w]|\bar B}\rangle  
                                                    \label{eq:ETC}
\end{equation}
and
\begin{eqnarray} 
M_{\rm S}(\bar B \rightarrow D^{[*]}\pi) &&                       
= -{i \over f_{\pi}}\Biggl\{\sum_n\Bigl({m_{D^{[*]}}^2 - m_B^2 
                                         \over m_n^2 - m_B^2}\Bigr)
  \langle{D^{[*]}|A_{\bar \pi}|n}\rangle
                         \langle{n|\tilde H_w|\bar B}\rangle  \nonumber \\
&& \qquad\qquad 
+ \sum_\ell\Bigl({m_{D^{[*]}}^2 - m_B^2 
                              \over m_\ell^2 - m_{D^{[*]}}^2}\Bigr)
\langle{D^{[*]}|\tilde H_w|\ell}\rangle
                  \langle{\ell|A_{\bar \pi}|\bar B}\rangle\Biggr\},  
                                                    \label{eq:SURF}
\end{eqnarray}
respectively, where $[V_\pi + A_\pi, \tilde H_w]=0$ has been used. 
(See Refs.\cite{hard pion} and \cite{suppl} for notations.) The 
equal-time commutator term, $M_{\rm ETC}$, has the same form as the
one in the old soft pion approximation but now has to be evaluated in 
the IMF. The surface term, $M_{\rm S}$, is given by a divergent of 
matrix element of $T$-product of axial vector current and 
$\tilde H_w$ taken between $\langle D^{[*]}|$ and $|\bar B\rangle$. 
However, in contrast with the soft pion 
approximation, contributions of single meson intermediate states can 
now survive when complete sets of energy eigen states are inserted 
between these two operators, and give a correction to the soft pion
approximation. (See, for more details, Refs.\cite{hard pion} and 
\cite{suppl}.) Therefore, $M_{\rm S}$ is given by a sum of all 
possible pole amplitudes, i.e., $n$ and $l$ in Eq.(\ref{eq:SURF}) 
run over all possible single meson states, not only ordinary 
$\{q\bar q\}$, but also hybrid $\{q\bar qg\}$, four-quark 
$\{qq\bar q\bar q\}$, glue-balls, etc. However, $n$ and $l$ as well 
as the external states are energy eigen states in the present case 
so that the states which sandwich $\tilde H_w$ should conserve
their spin in the rest frame. Since we consider Lorentz invariant
amplitudes, we should pick up $n$ and $l$ which conserve their 
spin\cite{Pakvasa}, although the amplitudes are now treated in the
IMF. Therefore, we drop, for example, vector meson pole contributions 
to the $u$-channel of pseudo scalar meson decays into two pseudo scalar 
meson states, although they have been taken into account for long 
time\cite{Marshak}. 

Since the $B$ meson mass $m_B$ is much higher than those of charm
mesons and since wave function overlappings between the ground-state 
$\{q\bar q\}_0$ and excited-state-meson states are expected to be
small, however, excited meson contributions will be small in these 
decays and can be safely neglected. Therefore the hard pion 
amplitudes as the non-factorizable long distance ones are 
approximately described in terms of 
{\it asymptotic ground-state-meson matrix elements} (matrix elements 
taken between single ground-state-meson states with infinite 
momentum) of $V_\pi$, $A_\pi$ and $\tilde H_w$. 

Amplitudes for dynamical processes of hadrons can be decomposed into 
({\it continuum contribution}) + ({\it  Born term}).  
Since $M_{\rm S}$ is given by a sum of pole amplitudes, 
$M_{\rm ETC}$ corresponds to the continuum 
contribution\cite{MATHUR} which can  develop a phase relative to
the Born term. Therefore we parameterize the ETC terms using
isospin eigen amplitudes and their phases. Since  the $D\pi$ final
state can have isospin $I={1 \over 2}$ and 
${3 \over 2}$, we decompose $M_{\rm ETC}$'s as  
\begin{eqnarray}
&&M_{\rm ETC}(\bar B^0\, \rightarrow D^+\pi^-) 
= \,\,\,\,\sqrt{{1 \over 3}}M_{\rm ETC}^{(3)}e^{i\tilde\delta_{3}} 
+ \sqrt{{2 \over 3}}M_{\rm ETC}^{(1)}e^{i\tilde\delta_{1}},  
                                               \label{eq:ETCMP}\\
&&M_{\rm ETC}(\bar B^0\, \rightarrow D^0\,\pi^0\,) 
= -\sqrt{{2 \over 3}}M_{\rm ETC}^{(3)}e^{i\tilde\delta_{3}} 
+  \sqrt{{1 \over 3}}M_{\rm ETC}^{(1)}e^{i\tilde\delta_{1}},  
                                                \label{eq:ETCZZ}\\
&&M_{\rm ETC}(B^- \rightarrow  D^0\,\pi^-) 
= \,\,\,\,\,\sqrt{3}M_{\rm ETC}^{(3)}e^{i\tilde\delta_{3}},            
                                                \label{eq:ETCZP}
\end{eqnarray}
where $M_{\rm ETC}^{(2I)}$'s are the isospin eigen amplitudes with
isospin $I$ and $\tilde\delta_{2I}$'s are the corresponding phase
shifts introduced. In the present approach, therefore, the final
state  interactions are included in the non-factorizable
amplitudes. This  is compatible with the fact that amplitudes with
final state interactions  are given by quark line diagrams which
belong to non-leading terms in the large $N_c$ expansion. 

Asymptotic matrix elements of $V_{\pi}$ and $A_{\pi}$  are
parameterized as 
\begin{eqnarray}
&&\langle{\pi^0|V_{\pi^+}|\pi^-}\rangle 
    = \sqrt{2}\langle{K^{+}|V_{\pi^+}|K^0}\rangle 
    = -\sqrt{2}\langle{D^{+}|V_{\pi^+}|D^0}\rangle 
    =  \sqrt{2}\langle{B^{+}|V_{\pi^+}|B^0}\rangle 
    = \cdots = \sqrt{2},                          \label{eq:MEV}\\
&&\langle{\rho^0|A_{\pi^+}|\pi^-}\rangle 
    = \sqrt{2}\langle{K^{*+}|A_{\pi^+}|K^0}\rangle 
    = -\sqrt{2}\langle{D^{*+}|A_{\pi^+}|D^0}\rangle 
    =  \sqrt{2}\langle{B^{*+}|A_{\pi^+}|B^0}\rangle 
    = \cdots = h,                                    \label{eq:MEA}
\end{eqnarray}
where $V_\pi$'s and $A_\pi$'s are isospin charges and their axial
counterpart, respectively. The above parameterization can be 
obtained by using asymptotic $SU_f(5)$ symmetry\cite{ASYMP}, or
$SU_f(5)$  extension of the nonet symmetry in $SU_f(3)$. Asymptotic
matrix  elements of $V_\pi$ between vector meson states can be
obtained by  exchanging pseudo scalar mesons for vector mesons with
corresponding  flavors in Eq.(\ref{eq:MEV}), for example, as 
$\pi^{0,-} \rightarrow \rho^{0,-}$, etc. 
The $SU_f(4)$ part of the above parameterization reproduces 
well\cite{suppl,Takasugi} the observed values of decay rates, 
$\Gamma(D^*\rightarrow D\pi)$ and $\Gamma(D^*\rightarrow D\gamma)$. 

In this way we can describe the non-factorizable amplitudes for the 
$\bar B \rightarrow D\pi$ decays as 
\begin{eqnarray}
&&M_{\rm NF}(\bar B^0\, \rightarrow D^+\pi^-) 
\simeq -i{\langle{D^0|\tilde H_w|\bar B^0}\rangle \over f_\pi}
\Biggl\{
\Biggl[{4\over 3}e^{i\tilde\delta_1} 
                        -{1\over 3}e^{i\tilde\delta_3}\Biggr]
                            +  \cdots  \Biggr\},  \label{eq:LMP}
\\
&&M_{\rm NF}(\bar B^0\, \rightarrow D^0\,\pi^0\,) 
\simeq -i{\langle{D^0|\tilde H_w|\bar B^0}\rangle \over f_\pi}
\Biggl\{
{\sqrt{2} \over 3}\Biggl[2e^{i\tilde\delta_1} 
                                  + e^{i\tilde\delta_3}\Biggr]
                               + \cdots  \Biggr\},  \label{eq:LZZ}
\\
&&M_{\rm NF}(B^- \rightarrow D^0\,\pi^-) 
\simeq\,\,\,\, 
       i{\langle{D^{0}|\tilde H_w|\bar B^0}\rangle \over f_\pi}
                                       \Biggl\{e^{i\tilde\delta_3} 
                                 +\cdots\Biggr\},   \label{eq:LZP}
\end{eqnarray}
where the ellipses denote the neglected pole contributions. 

In the case of the $\bar B \rightarrow D^*\pi$ decays, the matrix
element ${\langle V|\tilde H_w|P\rangle}$ should vanish because of
conservation of spin so that $M_{ETC}(\bar B \rightarrow D^*\pi)$ 
also should vanish but now $D$ and $B^*$ poles in the $s$- and
$u$-channels, respectively, survive, i.e., 
\begin{eqnarray}
&&M_{\rm NF}(\bar B^0\, \rightarrow D^{*+}\pi^-) 
\simeq {i\over f_\pi}
\langle{D^{0}|\tilde H_w|\bar B^{0}}\rangle 
\Biggl({m_{B}^2-m_{D^*}^2 \over m_{B}^2-m_{D}^2}\Biggr)               
          \sqrt{1\over 2}h + \cdots ,              \label{eq:LSMP}
\\
&&M_{\rm NF}(\bar B^0\, \rightarrow D^{*0}\,\pi^0\,) 
\simeq {i \over \sqrt{2}f_\pi}
\Biggl[\langle{D^{0}|\tilde H_w|\bar B^0}\rangle 
         \Biggl({m_{B}^2-m_{D^*}^2 \over m_{B}^2-m_{D}^2}\Biggr)
\nonumber \\
&&\hspace{60mm}+ \langle{D^{*0}|\tilde H_w|\bar B^{*0}}\rangle 
\Biggl({m_{B}^2-m_{D^*}^2 \over m_{B^*}^2-m_{D^*}^2}\Biggr)\Biggr]
              \sqrt{1\over 2}h +\cdots,  \label{eq:LSZZ}\\
&&M_{\rm NF}(B^- \rightarrow D^{*0}\,\pi^-) 
\simeq -{i \over f_\pi}\langle{D^{*0}|\tilde H_w|\bar B^{*0}}\rangle
\Biggl({m_{B}^2-m_{D^*}^2 \over m_{B^*}^2-m_{D^*}^2}\Biggr)
           \sqrt{1\over 2}h +\cdots,   \label{eq:LSZP}
\end{eqnarray}
where the ellipses denote the neglected excited meson contributions. 
Therefore the non-factorizable amplitudes in the hard pion 
approximation are controlled by the asymptotic ground-state-meson 
matrix elements of $\tilde H_w$ (and the possible phases). 

Now we evaluate the amplitudes given above. The factorized amplitudes 
in Table~I contain many parameters which have not been measured by 
experiments, {\it i.e.}, form factors, $F^{DB}_0(q^2)$,
$A^{D^*B}_0(q^2)$, $F^{\pi B}_1(q^2)$, etc., and decay constants, $f_D$,
$f_D^*$, $f_B$, etc.  The form factors $F_0^{DB}(0)$ and $A_0^{D^*B}(0)$
can be calculated by using the heavy quark effective theory
(HQET)~\cite{HQET}. The other form factors are concerned with light 
mesons and therefore have to be estimated by some other models. In color
favored decays, main parts of the factorized amplitudes depend on the
form  factor, $F_0^{DB}(0)$ or $A_0^{D^*B}(0)$, and the other form 
factors are included in minor terms proportional to $a_2$. Therefore our
result may not be lead to serious uncertainties although some model
dependent values of the  form factors are taken. (We will take, later, 
the values given in  Ref.\cite{Kamal}.) In the color suppressed $\bar
B^0\rightarrow D\pi^0$ and $D^{*0}\pi^0$ decays, however, the factorized
amplitudes are  proportional to the form factors, $F_0^{\pi B}(m_D^2)$ 
and $F_1^{\pi B}(m_{D^*}^{2})$, respectively. Since their values are 
model dependent, the result on the color suppressed decays may be a
little ambiguous, if non-factorizable contribution is less important. 
For the decay constants of heavy mesons, we assume $f_D \simeq f_{D^*}$
(and $f_B \simeq f_{B^*}$) since $D$ and $D^*$ ($B$ and $B^*$) are 
expected to be degenerate because of heavy quark symmetry~\cite{HQET} and
are approximately degenerate in  reality. Here we take 
$f_{D^*} \simeq f_D \simeq 211$ MeV and 
$f_{B^*} \simeq f_B \simeq 179$ MeV 
from a recent result of lattice QCD\cite{Lattice}. 
In this way, we can obtain the factorized amplitudes in the second 
column of Table~II, where we have neglected very small annihilation 
terms in the $\bar B^0 \rightarrow D^0\pi^0$ and $D^{*0}\pi^0$ 
decay amplitudes. 

To evaluate the non-factorizable amplitudes, we need to know the 
size of the asymptotic matrix elements of $\tilde H_w$ and $A_\pi$ 
taken between heavy meson states. The latter which was 
parameterized in Eq.(\ref{eq:MEA}) is estimated to be 
$|h|\simeq 1.0$\cite{hard pion,suppl} by using partially conserved 
axial-vector current (PCAC) and the observed rate\cite{PDG98}, 
$\Gamma(\rho \rightarrow \pi\pi)_{expt} \simeq 150$ MeV. 
For the asymptotic matrix elements, 
$\langle D^0|\tilde H_w|\bar B^0 \rangle$ and 
$\langle D^{*0}|\tilde H_w|\bar B^{*0} \rangle$, 
we treat them as unknown parameters and search phenomenologically for
their values to reproduce the observed rates for the 
$\bar B\rightarrow D^{[*]}\pi$ decays. To this, we 
parameterize these matrix elements using factorizable ones of 
$H_w^{BSW}$ as 
$\langle {D^{[*]0}|\tilde H_w|\bar B^{[*]0}} \rangle 
=B_H \langle {D^{[*]0}|H_w^{BSW}|\bar B^{[*]0}} \rangle _{FA}$ 
where $B_H$ is a parameter introduced and, for example, 
\begin{equation}
\langle {D^{0}|H_w^{BSW}|\bar B^{0}} \rangle _{FA} 
={G_F\over \sqrt{2}}V_{cb}V_{ud}
\Bigl({m_D^2 + m_B^2 \over 2}\Bigr) f_Df_Ba_2. 
\end{equation}
In this way, we obtain the hard 
pion amplitudes as the non-factorizable contributions listed in the 
third column of Table~II, where the CKM matrix elements have been 
factored out. 
\newpage
\begin{center}
\begin{quote}
{Table~II. Factorized and non-factorizable amplitudes for the 
$\bar B \rightarrow D\pi$ and $D^*\pi$ decays.
The CKM matrix elements are factored out. }
\end{quote}
\vspace{0.5cm}

\begin{tabular}
{l|l|l}
\hline
$\quad\,\,${\rm Decay}
&$\quad A_{\rm FA}\,(\times 10^{-5}$  GeV)
&$\qquad\qquad A_{\rm NF}\,(\times 10^{-5}$ GeV) 
\\
\hline
$\bar B^0 \rightarrow D^{+}\pi^-$
& $\quad 1.54\,a_1\Bigl\{
                  1 - 0.11\Bigl({a_2 \over a_1}\Bigr)\Bigr\}$
&$\, -3.70a_2B_H\,\Bigl\{
\Bigl[{4\over 3}e^{i\tilde\delta_1} 
                   -{1\over 3}e^{i\tilde\delta_3}\Bigr] 
                                              \Bigr\}$
\\
$\bar B^0 \rightarrow D^{0}\pi^0$
& $\, -1.29\,a_2\Bigl\{
                         {f_D \over 0.211\,\,{\rm GeV}}\Bigr\}$
& $\, -3.70a_2B_H\,\Bigl\{
{\sqrt{2} \over 3}\Bigl[2e^{i\tilde\delta_1} 
                                   + e^{i\tilde\delta_3}\Bigr]
                                              \Bigr\}$
\\
$B^- \rightarrow D^0\pi^-$
& $\quad 1.54\,a_1\Bigl\{
                   1 + 1.18\Bigl({a_2 \over a_1}\Bigr)\Bigr\}$ 
& $\quad 3.70a_2B_H\,\Bigl\{e^{i\tilde\delta_3}\Bigr\}$
\\
$\bar B^0 \rightarrow D^{*+}\pi^-$
& $\, -1.53\,a_1\Bigl\{
                    1 - 0.28\Bigl({a_2 \over a_1}\Bigr)\Bigr\}$
& $\quad 3.70a_2B_H\,\Bigl\{ 
- 0.694 \Bigr\}$
\\
$\bar B^0 \rightarrow D^{*0}\pi^0$
& $\quad 1.10\,a_2\Bigl\{{f_{D^*} \over 0.211\,\,{\rm GeV}}\Bigr\}$
& $\quad 3.70a_2B_H\,\Bigl\{
+ 0.00135 \Bigr\}$
\\
$B^- \rightarrow D^{*0}\pi^-$
& $\, -1.53\,a_1\Bigl\{
                 1 + 1.02\Bigl({a_2 \over a_1}\Bigr)\Bigr\}$ 
& $\quad\, 3.70a_2B_H\,\Bigl\{
-0.696 \Bigr\}$
\\
\hline
\end{tabular}

\end{center}
\vspace{0.5cm}

We now compare our result on the branching ratios, 
$B(\bar B\rightarrow D\pi)$ and $B(\bar B \rightarrow D^*\pi)$, 
with experiments, taking a sum of the factorized amplitude (the 
second column in the Table~II) and the non-factorizable amplitude 
(the third column in Table~II) as the total one. To this, we 
determine values of parameters involved. We take $V_{cb}=0.0395$ from 
the updated value $|V_{cb}|=0.0395 \pm 0.0017$\cite{PDG98}. 
For the coefficients $a_1$ and $a_2$ in $H_w^{BSW}$, we do not know 
their true values. According to Ref.\cite{BURAS}, NLO corrections to 
$a_1$ are small while corresponding corrections to $a_2$ may be not  
much smaller compared with the LO corrections and depend strongly on
the renormalization scheme. Therefore, we expect that the value,
$a_1=1.024$, with the LO corrections\cite{BURAS} is not very far 
from the true value and we take conservatively the above value of
$a_1$. For $a_2$, however, we consider two cases. We take $a_2=0.125$ 
with the LO QCD corrections~\cite{BURAS} as the case (i) and 
then we treat it as an adjustable parameter around the above 
$a_2=0.125$ as the case (ii). 
For the phases $\tilde\delta_1$ and $\tilde\delta_3$ arising from 
contributions of non-resonant multi-hadron intermediate states into 
isospin $I={1\over 2}$ and ${3\over 2}$ final states, they are 
restricted in the region $|\tilde\delta_{2I}| < 90^\circ$ since 
resonant contributions have already been extracted as pole amplitudes 
in $M_{\rm S}$ although their contributions are neglected as 
discussed before. For $B_H$, we here treat it as a free parameter. 

We now search for values of parameters, $\tilde\delta_1$,
$\tilde\delta_3$ and $B_H$ in the case (i), and $a_2$, 
$\tilde\delta_1$, $\tilde\delta_3$ and $B_H$ in the case (ii), to 
reproduce the phenomenologically estimated branching ratios (from the 
observed ones) for the $\bar B\rightarrow D^{[*]}\pi$ decays.  Large 
$\tilde\delta_1$, ($90^\circ > \tilde\delta_1\gtrsim 60^\circ$), and 
small $|\tilde\delta_3|$ are favored but our result is not very 
sensitive to the latter. For the $B_H$ parameter, $B_H\simeq 0.40$ in 
(i) but smaller values, $0.2 \gtrsim B_H \gtrsim 0.1$, in (ii) are 
favored. We list our results on the branching ratios in  
(i) $a_1=1.024$, $a_2=0.125$, $\tilde\delta_1=85^\circ$, 
$\tilde\delta_3=-5^\circ$ and $B_H=0.40$, 
and 
(ii) $a_1=1.024$, $a_2=0.19$, $\tilde\delta_1=85^\circ$, 
$\tilde\delta_3=-5^\circ$ and $B_H=0.15$ 
in Table~III, where we have used the central values, 
$V_{cb}=0.0395$, $V_{ud}=0.98$, 
$\tau(B^-) = 1.65\times10^{-12}$~s and 
$\tau(\bar B^0) = 1.56\times10^{-12}$~s, 
of their experimental data. $B_{\rm FA}$ and $B_{\rm tot}$ are given 
by the factorized 
amplitude and a sum of the factorized and 
non-factorizable ones, respectively. Values of $B_{\rm ph}$ 
have been obtained phenomenologically from $B_{\rm expt}$\cite{PDG98} 
in Sec.~I. $B_{\rm FA}$, in which the non-factorizable contributions 
are neglected, can reproduce fairly well the existing data. However, 
if we add the non-factorizable contributions, we can improve the fit 
to the phenomenologically estimated $B_{\rm ph}$ in both cases, (i) 
and (ii). It is seen that the non-factorizable contributions to the 
color favored $\bar B \rightarrow D\pi$ and $D^*\pi$ decays are 
rather small but still can interfere efficiently with the main 
amplitude given by the naive factorization. 
\newpage
\begin{quote}
{Table~III. Branching ratios ($\%$) for $\bar B \rightarrow D\pi$ 
and $D^*\pi$ decays where the central values of experimental 
data\cite{PDG98}, 
$V_{cb}=0.0395$, $V_{ud}=0.98$, 
$\tau(B^-) = 1.65\times10^{-12}$~s and 
$\tau(\bar B^0) = 1.56\times10^{-12}$~s, 
have been used. $a_2=0.125$ with the LO QCD corrections and 
$B_H=0.40$ in (i) and phenomenological $a_2=0.19$ and $B_H=0.15$ in 
(ii) have been taken, respectively, but 
$a_1=1.024$, $\tilde\delta_1=85^\circ$, $\tilde\delta_3=-5^\circ$ 
in both cases. $B_{\rm FA}$ and $B_{\rm tot}$ are given by the 
factorized amplitude and a sum of the factorized and non-factorizable 
ones, respectively. The values of phenomenologically estimated 
$B_{\rm ph}$ have been given in the text. 
} 
\end{quote}

\begin{center}
\begin{tabular}
{|c|c|l|l|c|c|}
\hline
\multicolumn{2}{|c|}
 {Decays} 
&\quad $B_{\rm FA}$ \quad
&\quad $B_{\rm tot}$ \quad
&\quad $B_{\rm ph}$ \quad
&\quad $B_{\rm expt}$ (*) \qquad
\\
\hline
{$B(\bar B^0\rightarrow D^+\pi^-)$}
&\begin{tabular}{c}
(i)\\
(ii)
\end{tabular}
&\begin{tabular}{c}
\hspace{2mm}{0.28}\\
\hspace{2mm}{0.27}
\end{tabular}
&\begin{tabular}{c}
\hspace{2mm}{0.30}\\
\hspace{2mm}{0.28}
\end{tabular}
&$0.28 - 0.34$
&$0.30 \pm 0.04$
\\
\hline
{$B(\bar B^0\rightarrow D^0\pi^0)$}
&\begin{tabular}{c}
(i)\\
(ii)
\end{tabular}
&\begin{tabular}{c}
\hspace{2mm}{0.003}\\
\hspace{2mm}{0.007}
\end{tabular}
&\begin{tabular}{c}
\hspace{2mm}{0.011}\\
\hspace{2mm}{0.012}
\end{tabular}
&\hspace{2mm}$0.006 - 0.012$\hspace{2mm}
&$< 0.012$
\\
\hline 
{$B(B^-\rightarrow D^0\pi^-)$}
&\begin{tabular}{c}
(i)\\
(ii)
\end{tabular}
&\begin{tabular}{c}
\hspace{2mm}{0.40}\\
\hspace{2mm}{0.46}
\end{tabular}
&\begin{tabular}{c}
\hspace{2mm}{0.49}\\
\hspace{2mm}{0.51}
\end{tabular}
&$0.53 \pm 0.05$
&$0.53 \pm 0.05$
\\
\hline
{$B(\bar B^0\rightarrow D^{*+}\pi^-)$}
&\begin{tabular}{c}
(i)\\
(ii)
\end{tabular}
&\begin{tabular}{c}
\hspace{2mm}{0.26}\\
\hspace{2mm}{0.25}
\end{tabular}
&\begin{tabular}{c}
\hspace{2mm}{0.30}\\
\hspace{2mm}{0.27}
\end{tabular}
&\hspace{2mm}$0.276 \pm 0.021$
&\hspace{2mm}$0.276 \pm 0.021$
\\
\hline
{$B(\bar B^0\rightarrow D^{*0}\pi^0)$}
&\begin{tabular}{c}
(i)\\
(ii)
\end{tabular}
&\begin{tabular}{c}
\hspace{2mm}{0.002}\\
\hspace{2mm}{0.005}
\end{tabular}
&\begin{tabular}{c}
\hspace{2mm}{0.002}\\
\hspace{2mm}{0.005}
\end{tabular}
&\hspace{2mm}$0.004 - 0.044$
&$< 0.044$
\\
\hline
{$B(B^-\rightarrow D^{*0}\pi^-)$}
&\begin{tabular}{c}
(i)\\
(ii)
\end{tabular}
&\begin{tabular}{c}
\hspace{2mm}{0.37}\\
\hspace{2mm}{0.42}
\end{tabular}
&\begin{tabular}{c}
\hspace{2mm}{0.43}\\
\hspace{2mm}{0.45}
\end{tabular}
&\hspace{2mm}$0.46 \pm 0.04$
&\hspace{2mm}$0.46 \pm 0.04$
\\
\hline
\end{tabular}

\end{center}
\vspace{0.5 cm}

\section{$\bar B \rightarrow J/\psi\bar K$ and $J/\psi\pi$ decays}

Now we study Cabibbo-angle favored $\bar B \rightarrow J/\psi\bar K$ 
and suppressed $B^- \rightarrow J/\psi\pi^-$ decays in the same way 
as in the previous section. Both of them are color suppressed and 
their kinematical condition is much different from the color favored 
$\bar B \rightarrow D\pi$ and $D^*\pi$ decays at the level of 
underlying quarks, {\it i.e.}, $b \rightarrow (c\bar c)_1\,+\,s$ in
the former but $b \rightarrow c\,+\,(\bar ud)_1$ in the latter. 
Therefore, dominance of factorized amplitudes in the 
$\bar B \rightarrow J/\psi\bar K$ and $B^- \rightarrow J/\psi\pi^-$ 
decays has no theoretical support and hence non-factorizable long 
distance contribution may be important in these decays. 

The factorized amplitude for the 
$\bar B \rightarrow J/\psi\bar K$ decays is given by 
\begin{equation}
M_{\rm FA}(\bar B \rightarrow J/\psi\bar K)
=-iV_{cb}V_{cs}
\Bigl\{
{G_F \over \sqrt{2}}a_2f_\psi F_1^{KB}(m_\psi^2)
\Bigr\}2m_\psi\epsilon^*(p')\cdot p. 
\end{equation}
The value of the decay constant of $J/\psi$ is estimated to be 
$f_\psi \simeq 380$~MeV from the observed rate\cite{PDG98} for the 
$J/\psi \rightarrow \ell^+\ell^-$. The value of the CKM matrix 
element $V_{cs}$ is given by $V_{cs} \simeq V_{ud} \simeq 0.98$. 
The value of the form factor $F_1^{KB}(m_\psi^2)$ has not been 
measured and its theoretical estimates are model dependent. 
We pick out tentatively the values of $F_1^{KB}(m_\psi^2)$ based on 
the following five models, {\it i.e.}, BSW\cite{BSW}, GKP\cite{GKP}, 
CDDFGN\cite{CDDFGN}, AW\cite{AW} and ISGW\cite{ISGW}, and list the 
corresponding $B_{\rm FA}(\bar B \rightarrow J/\psi\bar K)$ in 
Table~IV, where we have used  
$V_{cb}=0.0395$, 
$\tau_B^- = 1.65\times 10^{-12}$~s, 
$\tau_{\bar B^0} = 1.56\times 10^{-12}$~s  
as before. For $a_2$, we consider again two cases, i.e., 
\newpage 
\begin{quote}
{Table~IV. Branching ratios ($\%$) for the 
$\bar B \rightarrow J/\psi\bar K$ decays where the values of 
$F_1^{KB}(m_\psi^2)$ estimated in the five models, BSW, GKP, CDDFGN, 
AW and ISGW, in Refs.\cite{BSW}, \cite{GKP}, \cite{CDDFGN}, 
\cite{AW} and \cite{ISGW}, respectively, are used. Values of the other 
parameters involved are the same as in Table~III, where  $B_H'=B_H$
has been assumed. The data values are taken from Ref.\cite{PDG98}.}
\end{quote}

\begin{center}
\begin{tabular}
{|c|c|c|c|c|c|c|}
\hline
\multicolumn{2}{|c|}
{\hspace{3mm}Models}
& $\quad$BSW$\quad$
& $\quad$GKP$\quad$
& $\quad$CDDFGN$\quad$
& AW
& $\quad$ISGW$\quad$ 
\\
\hline
\multicolumn{2}{|l|}
{$\quad F_1^{K B}(m_\psi^2)\quad$}
&$\quad$0.565$\quad$
&$\quad$0.837$\quad$
&$\quad$0.726$\quad$
&$\quad$0.542$\quad$
&$\quad$0.548$\quad$
\\
\hline
{$\quad B_{\rm FA} \quad$} 
&\begin{tabular}{c}
(i)\\
(ii)
\end{tabular}
&{\begin{tabular} {c}
{0.015} \\
0.034
\end{tabular}
}
&{\begin{tabular} {c}
{$\quad$0.032$\quad$} \\
{$\quad$0.075$\quad$}
\end{tabular}
}
&{\begin{tabular} {c}
{$\quad$0.024$\quad$} \\
{$\quad$0.056$\quad$}
\end{tabular}
}
&{\begin{tabular} {c}
{$\quad$0.014$\quad$} \\
{$\quad$0.031$\quad$}
\end{tabular}
}
&{\begin{tabular} {c}
{$\quad$0.014$\quad$} \\
{$\quad$0.032$\quad$} 
\end{tabular}
}
\\
\hline
$\quad B_{\rm tot} \quad$
&\begin{tabular}{c}
(i)\\
(ii)
\end{tabular}
&
{\begin{tabular} {c}
{$\quad$0.040$\quad$} \\
{$\quad$0.052$\quad$}
\end{tabular}
}
&
{\begin{tabular} {c}
{$\quad$0.066$\quad$}\\
{$\quad$0.101$\quad$}
\end{tabular}
}
&
{\begin{tabular} {c}
{$\quad$0.055$\quad$}\\
{$\quad$0.079$\quad$}
\end{tabular}
}
&
{\begin{tabular} {c}
{$\quad$0.038$\quad$}\\
{$\quad$0.049$\quad$}
\end{tabular}
}
&
{\begin{tabular} {c}
{$\quad$0.038$\quad$}\\
{$\quad$0.050$\quad$}
\end{tabular}
}
\\
\hline
\multicolumn{2}{|l|}
{$\quad$Experiment$\quad$}
&\multicolumn{5}{|c|}
{\begin{tabular}{c}
$B(B^-\rightarrow J/\psi K^-) = (0.099 \pm 0.010)\,\,\%$ 
\\
$B(\bar B^0\, \rightarrow J/\psi\bar K^0\,) 
= (0.089 \pm 0.012)\,\,\%$
\end{tabular}
}
\\
\hline
\end{tabular}

\end{center}
\vspace{0.5 cm}
(i) $a_2=0.125$ and (ii) $a_2=0.19$. The results ($B_{\rm FA}$) from 
the factorized amplitudes for the values of $F_1^{KB}(m_\psi^2)$ 
listed in Table~IV are considerably smaller (except for the 
GKP\cite{GKP}) than the observations\cite{PDG98}, 
\begin{eqnarray}
&&B(B^- \rightarrow J/\psi K^-)_{\rm expt} 
= (0.099 \pm 0.010) \,\,\% \nonumber \\
&&B(\bar B^0 \,\rightarrow J/\psi\bar K^0\,)_{\rm expt} 
= (0.089 \pm 0.012) \,\,\%. \label{eq:psi-K}
\end{eqnarray}

Non-factorizable contributions to these decays are estimated by using 
a hard kaon approximation which is a simple extension of the hard 
pion technique in the previous section. With this approximation and
isospin symmetry, non-factorizable amplitude for the 
$\bar B \rightarrow J/\psi\bar K$ decays is given by 
\begin{equation}
M_{\rm NF}(\bar B \rightarrow J/\psi\bar K) 
= {i \over f_K}
{\langle \psi|H_w|\bar B_s^{*0} \rangle }
\Biggl({m_B^2 - m_\psi^2 \over m_{B_s^*}^2 - m_\psi^2}\Biggr)
\sqrt{1 \over 2}h \, +\, \cdots, 
\end{equation}
where the ellipsis denotes neglected contributions of excited 
mesons\cite{Close} and 
$\langle \bar B_s^0|V_{K^+}|B^- \rangle = -1$ and 
$\sqrt{2}\langle \bar B_s^{*0}|A_{K^+}|B^- \rangle = -h$ 
which are flavor $SU_f(3)$ extensions of Eqs.(\ref{eq:MEV}) and 
(\ref{eq:MEA}) have been used. Asymptotic matrix element, 
$\langle \psi|\tilde H_w|\bar B_s^{*0} \rangle$, is parameterized in 
the same way as $\langle D^{*0}|\tilde H_w|\bar B^{*0} \rangle$
before. Then the total amplitude for the 
$\bar B \rightarrow J/\psi\bar K$ decays is approximately given by 
\begin{equation}
M_{\rm tot}(\bar B \rightarrow J/\psi\bar K) 
\simeq -iV_{cb}V_{cs}\bigl\{5.73F_1^{KB}(m_\psi^2)\,+\,5.16B_H'
\bigr\}a_2\times10^{-5}\,\, {\rm GeV}         \label{eq:amp-psi-k}
\end{equation}
where $f_K \simeq 160$~MeV and 
$f_{B^*_s} \simeq f_{B_s} \simeq 204$~MeV from the updated lattice 
QCD result\cite{Lattice} have been taken. $B_H'$ is a parameter 
corresponding to $B_H$, i.e., 
\begin{equation}
\langle \psi|\tilde H_w|\bar B_s^{*0} \rangle 
= B_H'\langle \psi| H_w^{BSW}|\bar B_s^{*0} \rangle_{\rm FA} .
\end{equation}

When we take $a_2=0.19$ and $B_H'=0.15$ as before, we can reproduce 
considerably well the existing experimental data on the 
$\bar B \rightarrow J/\psi\bar K$ decays by $B_{\rm tot}$ although 
the result depends sharply on the values of the form factor 
$F_1^{KB}(m_\psi^2)$. 

For the Cabibbo-angle suppressed $B^- \rightarrow J/\psi\pi^-$, 
the same technique and values of parameters as the above lead to 
\begin{equation}
M_{\rm tot}(B^- \rightarrow J/\psi\pi^-) 
\simeq -iV_{cb}V_{cd}\bigl\{5.73F_1^{\pi B}(m_\psi^2)\,
                +\,5.46B_H'\bigr\}a_2\times10^{-5}\,\, {\rm GeV}.      
                                              \label{eq:amp-psi-pi}
\end{equation}
Using 
$F_1^{\pi B}(m_\psi^2) \simeq F_1^{KB}(m_\psi^2)$ 
expected from $SU_f(3)$ symmetry, we obtain 
\begin{equation}
B_{\rm tot}(B^- \rightarrow J/\psi\pi^-)\simeq 
\Biggl|{V_{cd} \over V_{cs}}\Biggr|^2
B_{\rm tot}(B^- \rightarrow J/\psi K^-)
\end{equation}
which is well satisfied by experiment\cite{PDG98}. 
$B_{\rm tot}(B^- \rightarrow J/\psi\pi^-)$ from the amplitude 
Eq.(\ref{eq:amp-psi-pi}) which includes both of the factorized 
amplitude and the non-factorizable one can reproduce the existing 
experimental data\cite{PDG98}, 
\begin{equation}
B(B^- \rightarrow J/\psi\pi^-)_{\rm expt} 
= (5.0\,\pm\,1.5)\times 10^{-5},
                                               \label{eq:psi-pi}
\end{equation}
by taking (i) $a_2=0.125$ and $B_H'=0.40$, and (ii) 
$a_2\simeq 0.19$ and $B_H'=0.15$ as before, although $B_H'=B_H$ is not 
necessarily required.

\section{Summary}

In summary, we have investigated the $\bar B \rightarrow D\pi$ and 
$D^*\pi$ and found that the existing data on their branching ratios
are not always compatible with each other, i.e., the r.h.s. of 
Eq.(\ref{eq:phase-diff}) is over unity for some values of 
$R_{00}^{[*]}$ and $R_{0-}^{[*]}$. Then we have obtained
phenomenologically allowed values of their branching ratios, 
$B_{\rm ph}$, which keep the r.h.s. of Eq.(\ref{eq:phase-diff}) 
approximately less than unity. Next, we have studied the 
$\bar B \rightarrow D\pi$, $D^*\pi$, $J/\psi \bar K$ and 
$J/\psi\pi^-$ decays describing their amplitude by a sum of 
factorizable and non-factorizable ones. The former amplitude has been 
estimated by using the naive factorization while the latter has been 
calculated by using a hard pion (or kaon) approximation in the
infinite momentum frame. 
The so-called final state interactions (corresponding to the NLO terms 
in the large $N_c$ expansion) have been included in the
non-factorizable long distance contributions. The non-factorizable 
contribution to the color favored $\bar B \rightarrow D\pi$ and 
$D^*\pi$ decays is rather small and therefore the final state 
interactions seem to be not very important in these decays although 
still not necessarily negligible. By taking $a_1\simeq 1.024$ with 
the LO QCD corrections and the phenomenological $a_2\simeq 0.19$ 
which has been suggested previously\cite{Kamal,CLEO}, the observed 
branching ratios for these decays can be well reproduced in terms of 
a sum of the hard pion amplitude and the factorized one. Namely, 
the factorized amplitudes are dominant but not complete and long 
distance hadron dynamics should be carefully taken into account in 
hadronic weak interactions of $B$ mesons. 

In color suppressed $\bar B^0 \rightarrow  D^0\pi^0$, 
$\bar B\rightarrow J/\psi \bar K$ and $J/\psi\pi^-$ decays, 
non-factorizable long distance contributions are more important. 
In particular, in the $\bar B \rightarrow J/\psi\bar K$ decay, long 
distance physics should be treated carefully. When $a_2\simeq 0.125$ 
with the LO QCD corrections is taken instead of the phenomenological 
$a_2\simeq 0.19$, it may be hard to reproduce the observed values of 
$B(\bar B \rightarrow J/\psi \bar K)$ and 
$B(B^- \rightarrow J/\psi\pi^-)$ 
even by taking a sum of factorized and non-factorizable amplitudes 
as long as $B_H'=B_H\simeq 0.4$ is taken. 

The non-factorizable amplitudes are proportional to asymptotic 
ground-state-meson matrix elements of $\tilde H_w$, i.e., 
$B_H$ or $B_H'$. To reproduce large rates for the color favored 
$\bar B \rightarrow D\pi$ and $D^*\pi$ decays, the non-factorizable
contributions are needed ($B_H \neq 0$) while too large values of
$B_H$ and $B_H'$ will lead to too large rates for the color suppressed
decays. However, their numerical results are still ambiguous since the
amplitudes for the color suppressed decays depend sharply on model
dependent form factors. 

Therefore more precise measurements of branching ratios for the color
suppressed decays, in particular, $B(\bar B \rightarrow D^0\pi^0)$, 
are useful to determine the non-factorizable long distance 
contributions in hadronic weak decays of $B$ mesons. 

\vskip 2cm

This work was supported in part by the Grant-in-Aid from the Ministry 
of Education, Science and Culture. 
The author would like to appreciate Prof. A.~I.~Sanda, 
Prof. Y.~Okada, Prof. Y.~Y.~Keum and Dr. D.~X.~Zhang  for comments. 
He also would like to thank Prof. S.~Pakvasa, Prof. H.~Yamamoto, 
Prof. S.~F.~Tuan and the other members of high energy physics group, 
University of Hawaii for their discussions, comments and hospitality 
during his stay there.


\begin{references}
\bibitem{hadron-dynamics} Dynamical contributions of hadrons play an 
important role in hadronic processes at low energies, for example, as 
pion exchange mechanism describes well nuclear force at long 
distances although its underlying dynamics must be QCD. 
In $B$ decays, such contributions will survive but their role may be 
less important since $m_B$ is large.
\end{references}
\end{document}